# On the relation between the microscopic structure and the sound velocity anomaly in elemental melts of groups IV, V, VI

Yaron Greenberg[1,2], Eyal Yahel[2], El'ad N. Caspi[2], Brigitte Beuneu[3], Moshe P. Dariel[1], Guy Makov[1]

[1] Department of Materials Engineering, Ben Gurion University of the Negev, Beer-Sheva 84105, Israel.

[2] Department of Physics, NRCN, P.O.Box 9001, Beer-Sheva 84190, Israel.

[3] Laboratoire Leon Brillouin (CEA-CNRS), CEA/Saclay, 91191 Gif-sur-Yvette Cedex, France

## Abstract

The sound velocity of some liquid elements of groups IV, V and VI, as reported in the literature, displays anomalous features that set them apart from other liquid metals. In an effort to determine a possible common origin of these anomalies, extensive neutron diffraction measurements of liquid Bi and Sb were carried out over a wide temperature range. The structure factors of liquid Sb and Bi were determined as a function of temperature. The structure of the two molten metals was carefully analyzed with respect to peak locations, widths and coordination number in their respective radial distribution function. The width of the peaks in the radial distribution function were not found to increase and even decreased within a certain temperature range. This anomalous temperature dependence of the peak widths correlates with the anomalous temperature dependence of the sound velocity. This correlation may be accounted for by increasing rigidity of the liquid structure with temperature. A phenomenological correlation between the peak width and the sound velocity is suggested for metallic melts and is found to agree with available data for normal and anomalous elemental liquids in groups IV-VI.

## Introduction

Simple elemental liquid metals display smooth, continuous and monotonous temperature dependence of their physical properties as well as of their microscopic structure above the melting point [1]. In contrast, many liquid metals of groups IV to VI exhibit an anomalous temperature dependence of the thermo-physical properties above their melting point. The anomalies include: (i) non-monotonous behavior of the sound velocity [2,3,4,5,6,7,8,9,10]; (ii) melting curve anomalies [11]; (iii) deviation from the universal temperature dependence of the configurational entropy [12]; and (iv), the presence of a complex structure in the pair correlation function, specifically, an additional feature on the high angle side of the first hard-sphere peak [13,14,15,16,17,18,19,20,21,22,23,24,25,26,27,28,29,30,31,32,33,34].

Focusing on the sound velocity in the liquid metallic pnictides, As, Sb and Bi, we see that in liquid arsenic (l-As) the sound velocity increases sharply with temperature above the melting point towards an as yet unidentified maximum [4]. In liquid antimony (l-Sb) the sound velocity increases slightly to a maximum at 1168K, 265 degrees above the melting point [6]. In liquid bismuth (l-Bi) the sound velocity is approximately temperature independent from the melting point to approximately 600K before declining non-linearly i.e., parabolically [6]. This anomalous behavior has not found yet a theoretical explanation.

The structure of the radial distribution function in the metallic pnictides exhibits an additional feature between the first and second hard sphere (HS) peaks, which manifests itself in an additional intermediate peak in As, as a hump in Sb and as a shoulder in Bi. The position of this additional feature (r) relative to the first HS peak ($R_1$) is approximately $r/R_1$=1.5 in l-As [25], and 1.4 in l-Sb [26] and l-Bi [30]. The position of the second HS peak ($R_2$) relative to the first peak is at approximately 2.5 in l-As and 2.1 in l-Sb and l-Bi. These results indicate that the distorted structure observed in l-As, diminishes along the metallic pnictide series and approaches a closed-packed structure, i.e., $r/R_1$~1.41 and $R_2/R_1$~2. In agreement with this interpretation, the number of nearest neighbors seems to increase along the series.

An attempt has been made to explain the anomalous structure of the pair correlation function of these liquids on the basis of their electronic structure. Specifically, these studies have pointed at a possible common origin for the anomalous structure in the pair correlation function, namely, the extension of structural distortions originating in the solid phase into the liquid. This was suggested for the group V elements**,** As, Sb and Bi that are known to form a rhombohedral A7 structure with three fold coordination symmetry in the solid phase and which originates from a Peierls distorted simple cubic lattice [25,35,36,37]. An alternative interpretation has also been advanced according to which, the structural distortion are attributed to Friedel oscillations [38].

Whether the Peierls distortion extends to the liquid state or not, is probably controlled by the ratio of the distortion energy to the thermal energy ($k_B$T). One can show by theoretical considerations that for the light group V elements, the distortion energy exceeds 0.1ev [37]. As a result, at the melting temperature, the entropy is not large enough to obliterate the Peierls distortion. A larger thermal energy (i.e., higher temperature) is required to overcome this distorted local arrangement. For the heavy elements, the distortion energy is expected to be smaller than the thermal energy, and therefore, the Peierls distortion is overcome at lower temperatures [37]. However, a relation between the electronic Peierls distortion of the ionic structure and the macroscopic thermodynamic anomalies has not yet been established.

Although, the physical properties and structures of liquid elemental metals have been studied extensively, much of the work has focused on the temperature range near the melting point [1]. Studies of the properties of liquid metals at higher temperatures are scarcer, primarily because of the experimental challenges involved in such work. It is of interest to explore in detail the correlation between the macroscopic thermodynamic properties and the microscopic structure at high temperatures. Since the features in the temperature dependence are relatively small and their variation is even smaller, a preliminary condition for such a comparison is the availability of high precision (<1%) observations. Much of the data previously published is not sufficiently precise for this purpose. Recently we have demonstrated the capability to perform high precision measurements of the sound velocity [6], density and structure factor in liquid metals [30].

In the present communication, we explore the evolution of the microscopic structure of l-Sb and l-Bi as a function of temperature by high precision neutron diffraction measurements. We find that the temperature variation of the structure can be correlated with the temperature dependence of the sound velocity. In particular the width of the peaks in the radial distribution function and the line-shape of g(r) exhibit the same temperature dependence as the sound velocity, suggesting a common microscopic origin. We discuss these findings in the context of the structure of Group IV-VI liquid metals.

## Experimental

Neutron diffraction measurements of liquid antimony and bismuth, were undertaken at the 7C2 two-axis diffractometer at Laboratoire Léon Brillouin (LLB) on the hot source of the reactor Orphée at Saclay [39]. A monochromatic neutron beam with $\lambda$=0.702Å was used in both cases. The q range available at the 7C2 diffractometer is 0.2-15.3$Å^{-1}$ . In this range data were collected at intervals of approximately 0.014$Å^{-1}$. Sample dimensions were 10 mm and 8 mm in diameter and 50 mm in length for antimony and bismuth respectively. The samples were mounted in a sealed amorphous silica capsule (10/12 mm, 8/10 mm inner/outer diameters respectively). The beam size was 12 mm wide and 50 mm in height. The silica capsule containing the sample was evacuated and filled with high purity He to approximately 300mmHg and was subsequently sealed. The temperature was measured by 3 thermocouples in contact with the bottom of the capsule. The vanadium furnace consisted of a vertical cylinder (30 mm diameter, 0.01 mm thickness and 300 mm height).

Measurements at 21 different temperature points in the range of 923-1423K were taken from a high purity (5N) antimony sample. Measurements at 14 different temperature points in the range of 553-1373K were taken from a high purity (5N) bismuth sample. Empty quartz capsule diffraction patterns were measured at several different temperature points in the relevant temperature ranges. These data were later used to subtract the background contribution in the analysis.

All neutron data analysis and some additional calculations, e.g. the calculating g(r) out of S(q) were performed using an analysis program developed at the LLB [40].

Conventional data reductions were applied for background, cell and furnace contributions, incoherent and multiple scattering [41,42,43]. Inelastic effects were treated by a Placzek type correction. In order to validate the absolute values of the static structure factor, S(q), a diffraction pattern was separately determined for a vanadium sample of the same dimensions as the sample.

## Results and Analysis

The static structure factor curves, S(q), measured for liquid antimony and for liquid bismuth are presented in Figure 1. The typical statistical error in S(q) is *ca.* 1% arising from the total number of counts. Radial distribution functions, g(r), were derived from these data, using the Fourier transform, and are presented in Figure 2. The typical error in g(r) is of the same magnitude.

An investigation of the temperature dependence of the g(r) curves was carried out, for both elements, by evaluating the height, width and position of the first and second hard sphere peaks. Throughout the entire temperature range, the positions of the first and second peaks in the radial distribution function in l-Sb were found to be constant at approximately 3.05Å and 6.30Å, respectively, within the estimated experimental error. The error in determining the peak position is determined by the statistical error in the peak heights and by the peak widths. Therefore the error varies between peaks, taking values of approximately ±0.02Å for the first peak and ±0.05Å for the second peak and an intermediate value of ca ±0.03Å for the hump. Unexpectedly, the height of both the first and the second peaks is observed to decrease only slightly with increasing temperature, by less than 5% over a range of 500 degrees (Figure 2). This behavior is in marked contrast to that of other metals such as tin, which declines by roughly 15% over a similar temperature range [20].

At the melting point of Sb, on the right hand side of the first peak in the radial distribution function, a hump is observed at approximately 4.15±0.03 Å as shown in Figure 2. The ratio of the radial position of this hump to that of the first peak is 1.36 and has been interpreted as indicating a distortion of a simple cubic-like structure of the liquid [27,44]. With increasing temperature, the radial position of the hump's maximum moves towards lower r values i.e. towards the main peak (Figure 3). In parallel, the hump's height with respect to the preceding minimum in g(r) diminishes gradually until it disappears completely above 1148K (Figure 3B). At some temperature between 1148K and 1173K the minimum between the main peak and the hump in the g(r) curve becomes imperceptible. Thus the hump transforms into a shoulder. We identify this disappearance of the minimum in g(r) as indicating a change in the liquid structure. From this temperature and above it is no longer possible to determine a position of the hump (Figure 3C), and we can only observe a gradual disappearance of the shoulder with temperature.

For l-Bi, throughout the whole measured temperature range, the positions of the first and second peaks are found to be constant, within the experimental error, at 3.26Å and 6.70Å, respectively, similar to the behavior of l-Sb. However, the heights of the first and second peaks in the g(r) curves decrease linearly by approximately 30% and 60%, respectively, over the measured 820 degrees temperature range (Figure 2). Near the melting temperature, at a radial distance of about 4.6Å, the presence of a shoulder on the first peak can be clearly observed. This observation is in distinction to the hump observed in l-Sb near the melting point. The position of the shoulder relative to that of the first peak is 1.41, also in agreement with the simple cubic-like interpretation [27,44]. The region of the shoulder in g(r) of l-Bi is shown in Figure 4. Note, that because of the presence of a shoulder and not a hump in the l-Bi data it is not possible to repeat the previous analysis preformed for l-Sb. The shoulder decreases as the temperature increases from the melting point (553K) and is barely visible at the high end of our measurement range.

The coordination number, N(r), is defined as the product of the atomic density and the volume weighted by the radial distribution function, $V(r_1,r_2)$:

$$(1) \quad N(r_1, r_2, T) = \rho(T)V(r_1, r_2, T) = 4\pi\rho(T)\int_{r_1}^{r_2} r^2 g(r,T)dr$$

where $r_1$ and $r_2$ are the radial limits of integration and T is the temperature. The density values, $\rho(T)$, were taken from previous studies for l-Sb [45] and for l-Bi [30]. Two regions of interest were *defined*: (a) main peak – from the exclusion distance ($r_1$) to the minimum/inflection point ($r_2$); (b) hump/shoulder – from the minimum/inflection point ($r_2$) to the minimum between the first and second HS peaks ($r_3$). The values selected by us for these parameters for both elements are given in Table I.

Table I : Radial parameters for nearest neighbors in liquid antimony and bismuth.

| | $r_1$[Å] | $r_2$ [Å] | $r_3$ [Å] |
|---|---|---|---|
| Sb | 2.50 | 3.80 | 5.10 |
| Bi | 2.65 | 3.95 | 5.40 |

An analysis of the temperature dependence of the coordination number, N(T), was carried out for l-Sb and l-Bi. The results are shown in Figure 5. The estimated error in determining N(T) is *ca.* 1.5%, which consists of two contributions: a statistical error in g(r) arising from S(q) (*ca.* 1%) and the error in determining the limits of integration (*ca.*

1%). The error in the density is an order of magnitude smaller, hence is negligible. For l-Sb, the number of atoms in the first shell decreases from 6.6±0.1 atoms at the melting point to 6.1±0.1 atoms at 1423K. The number of atoms in the shoulder decreases from 10.1±0.1 at melting to 9.5±0.1 atoms at the edge of the temperature range. It follows from Eq. (1) that there are two contributions to any change in the coordination number: a change in the density and a change in the volume. For the main peak we find that the volume weighted by the radial distribution function decreases by approximately 4% over the measured temperature range, whereas the density decreases by ca. 5%. Therefore the change in the coordination number arises both from a change in the density and a rearrangement of the liquid in comparable measures.

For l-Bi, one can observe that over the entire measured temperature range, the number of atoms in the first shell decreases by approximately one atom, from 8.0±0.1 atoms to 6.9 ±0.1 atoms. The number of atoms included in the shoulder decreases by less than one atom, from 10.3±0.1 atoms to 9.6±0.1 atoms. As discussed in previous work [30], we have shown that the change in coordination number is entirely determined by the change in the density with the total area under the radial distribution function remaining unchanged. However, a redistribution of the radial density at approximately 1100K points to a structural change of the liquid [30].

The peak structure in both l-Sb and l-Bi displays a strong asymmetry that results from the partial overlap between the hard-sphere peaks and the additional structure represented by the shoulder/hump. Therefore, we characterize the width of the peaks by the full width at half maximum (FWHM). The results of this analysis are presented in Figure 6. For "normal" metals the peak broadens with temperature due to the increase of the ratio of the thermal to structure energies in the liquid structure that leads to increased disorder [1]. For l-Sb, we find that the width of the first peak in the g(r) curves is practically temperature independent up to a temperature of approximately 1250K, from where on it increases. Even more surprisingly, the width of the second peak decreases with increasing temperature and exhibits a clear minimum in the 1150-1200K range. The temperature dependence of the widths of the first and second peaks in l-Bi is also presented in Figure 6. The FWHM increase for both peaks with increasing temperature throughout the measured temperature range and follows parabolic temperature dependence.

## Discussion

For both elements, our neutron diffraction data extend the measured temperature range and sample it more densely. For l-Bi it is in excellent agreement with previously published data [19,28,29,46,47,48,49]. For l-Sb, our data agree with previously published results, measured by x-ray and neutron diffraction [48]. We note that at 925K, a discrepancy can be observed when compared to other neutron diffraction data [50,51][§].

The coordination numbers found at the melting temperature for l-Bi were 8.0 in the first shell and 10.3 in the shoulder, in good agreement with previously published data [19,29,46]. The coordination numbers found for l-Sb were 6.6 for the first shell and 10.2 for the shoulder in good agreement with previously published data [13,27,44]. The coordination number in the metallic pnictides increases along the series from approximately 3 in As, through 6 in Sb [44] to 8 in Bi. This reflects the decrease of the Peierls distortion energy along the column in the periodic table.

In the temperature range studied we do not observe any transitions or transformations in the microscopic structure of l-Sb that correlate with the anomalous maximum in the sound velocity. However, we do find that the temperature dependence of the radial distribution function for l-Sb relative to l-Bi exhibits three significant features: (i) the height of the first peak in the radial distribution function is essentially temperature independent within the measured range as opposed to the observations in other metals e.g. in Bi where the peak height decreases by tens of percent within a similar temperature range. This implies that the probability to find an atom in the first shell does not decrease with temperature. One possible explanation is the existence of strong bonding in the first shell of l-Sb. A similar phenomenon can be observed in l-As under pressure. [27,44] (ii) Peak widths measured by FWHM display non-monotonous temperature dependence, as shown in Figure 6. In particular, the 1$^{st}$ peak has a constant width up to approximately 1200K at which point it begins to increase rapidly. We conclude that if we consider the thermal effect which broadens the peaks, there must be an equivalent counter-effect up to this temperature, possibly arising from bond strengthening. Even more dramatically, the 2$^{nd}$ peak width decreases with increasing temperature up to a minimum at approximately 1200K. In contrast, the FWHM of both the first and second peaks in Bi increase monotonously with temperature, following a parabolic trend. (iii) Finally, the hump on the right hand side of the first peak in l-Sb diminishes with temperature and transforms into a shoulder in the vicinity of 1200K. Thus, we see that there is consistent anomalous temperature dependence of the pair correlation function in l-Sb up to approximately 1200K and that this behavior can be interpreted as indicating exceptionally strong bonding in the melt. This conclusion is supported by a high pressure study of l-Sb which showed that the liquid structure is relatively insensitive to pressure up to about 5GPa indicative of strong bonding [27].

[§] Most surprisingly, our bismuth data stands in excellent agreement with the antimony data published in reference in [50,51], which raises the possibility of a typographical error.

High precision measurements of the temperature dependence of sound velocity in l-Sb and l-Bi were recently published [6]. A clear maximum was observed for l-Sb at 1168K. In l-Bi a plateau was observed at the melting point extending over a range of 60 degrees, followed by approximately parabolic behavior (Figure 6).

Based on these phenomenological observations, we propose the existence of a correlation between the temperature dependence of the velocity of sound and that of the structure in these melts. Such a correlation can be interpreted qualitatively as arising from the relation between the sound velocity and the compressibility. Thus, the existence of strong bonding in Sb causes the structure to retain relative rigidity, i.e., low compressibility. In parallel, the sound velocity maintains a non-decreasing constant value. In further agreement with this picture, the first peak height and width do not vary up to approximately 1200K.

This phenomenological correlation between the sound velocity and the temperature dependence of the structural rigidity is also exhibited by l-Bi, displaying a parabolic behavior for both properties. Further comparisons with additional metals are hampered by the scarcity of high quality data over a significant temperature range. A relatively good example is l-Sn [20] where the pair correlation function was measured at six temperature points. From these data we could only obtain the widths of the first peak which are shown in Figure 7. These data exhibit linear temperature dependence in agreement with our previously measured sound velocity curve [6]. In l-Te, the sound velocity increases rapidly up to a wide maximum at approximately 1075K [9,52]. From measurements of the pair correlation function reported in [32] we obtained the widths of the first peak and found that it decreases dramatically by approximately 30% from the melting point up to the temperature of the sound velocity maximum, as shown in Figure 7.

A hint of the microscopic mechanism behind these phenomena can be obtained by considering the significance of the transformation of the hump on the right hand side of the first peak in l-Sb into a shoulder as the temperature is increased. At low temperatures, the existence of a hump with a minimum in the radial distribution function between it and the first peak indicates that the occupation probability decreases non-monotonically as a function of the radial distance. As the temperature is increased the hump transforms into a shoulder and the minimum between the hump and the first peak disappears, i.e., the probability to find an atom as a function of the radial distance now decreases monotonically to the minimum between the first and second hard sphere peaks. In terms of the atomic configuration, we can interpret the presence of the hump as arising from the existence of two distinct atomic positions in the first shell with different radial distances and an energy barrier between them. As the temperature is increased, the barrier is overcome and the two atomic positions merge into a unified, albeit wide, state expressing itself in the radial distribution function by a broad first peak with a shoulder. The electronic origin of this barrier probably lies in the Peierls distortion mechanism. Support for this assumption is provided by the conclusion from electrical resistivity measurements [53] and EXAFS studies [26] that the Peierls barrier in l-Sb is overcome at approximately $800^0$C. Thus, in l-As, where the Peierls distortion is expected to be the largest, the hump is actually an independent peak. Future high temperature measurements may show that this peak diminishes into a shoulder with increasing temperature. In l-Bi where the distortion energy is the smallest, the hump is already a shoulder at the melting temperature. Possibly in a super-cooled state, a hump in the radial distribution function could be observed as well as a maximum in the sound velocity.

## Conclusions

To summarize, we have performed high precision measurements of the structure factor of l-Sb and l-Bi, by densely sampling a wide temperature range. We find that the temperature dependence of the peak widths in the radial distribution function of l-Sb is anomalous up to roughly 1200K, a temperature which corresponds to the maximum in the sound velocity curve. We introduce a phenomenological correlation between the peak widths in the radial distribution function and the sound velocity. This correlation has its origin in the underlying microscopic structure, which becomes more rigid with increasing temperature up to the sound velocity maximum at which point it reverts to normal softening with temperature. We therefore portray the melt as a system undergoing two counteracting processes in which the thermal motion and the rigidity of the first two shells play a role. This correlation is also observed between our sound velocity and liquid structure measurements in Bi and in published data of As, Sn and Te.

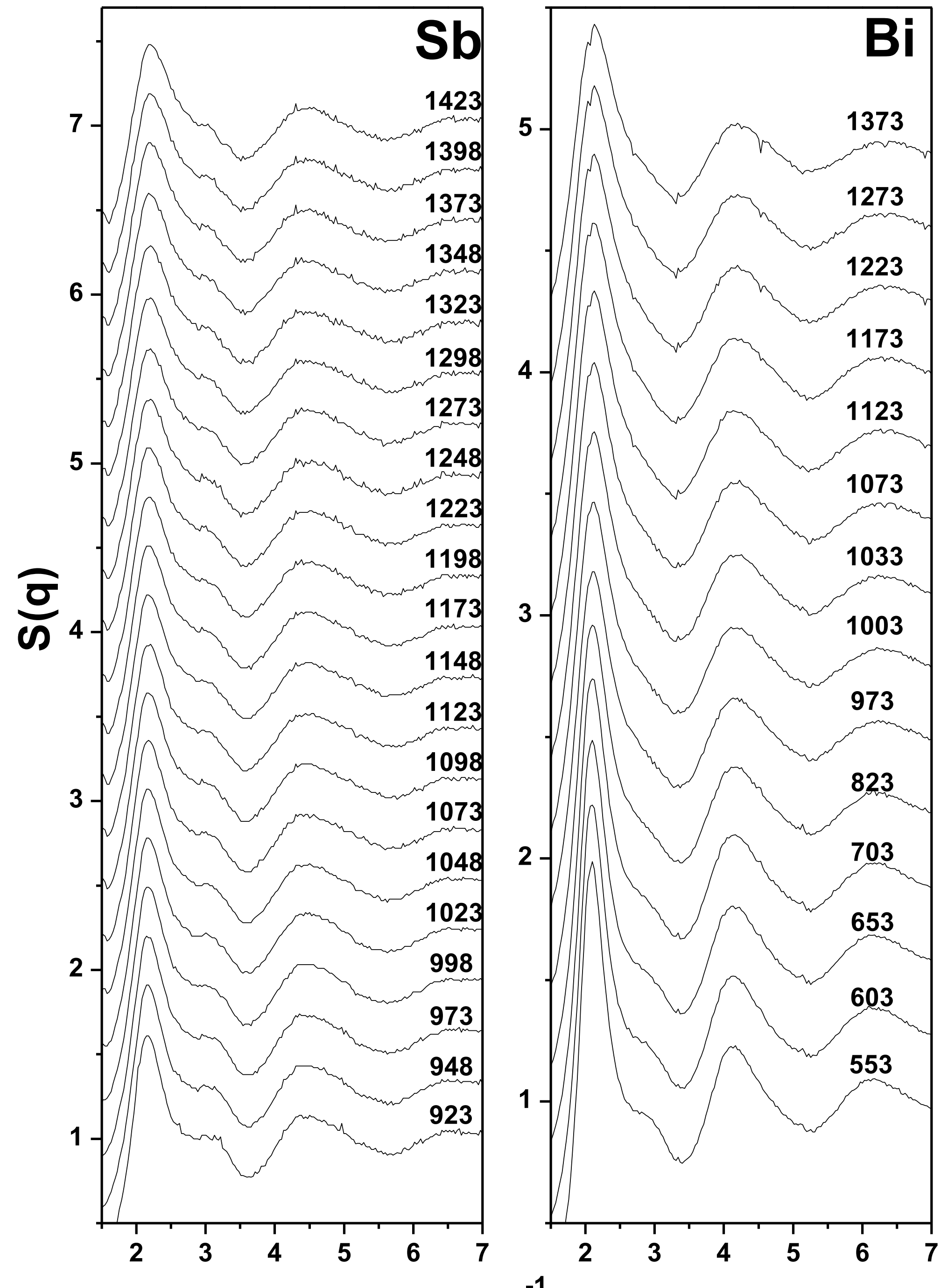

Figure 1: The structure factor curves at various temperatures, for liquid antimony and liquid bismuth.

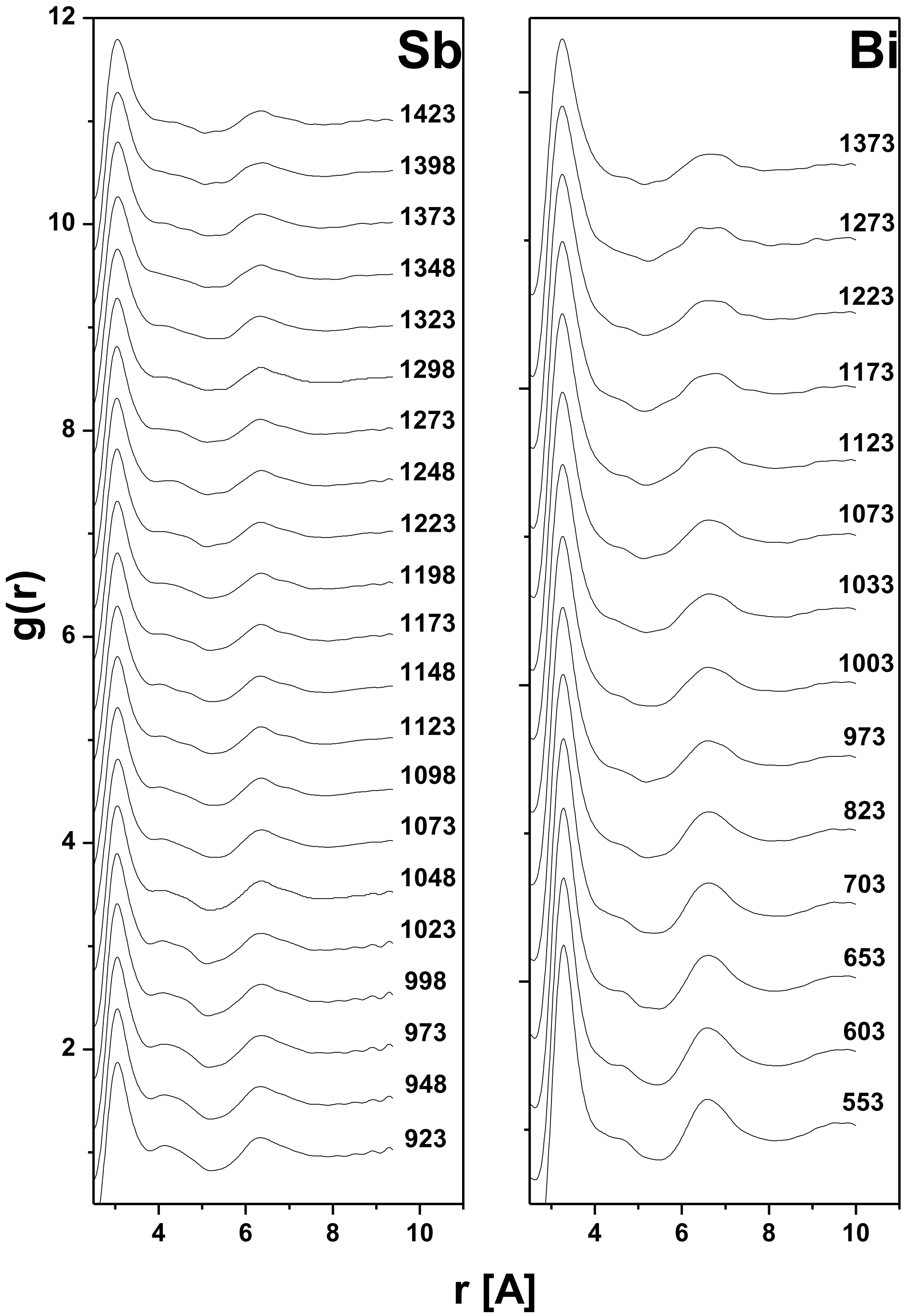

Figure 2: The radial distribution function curves at various temperatures, for liquid antimony and liquid bismuth.

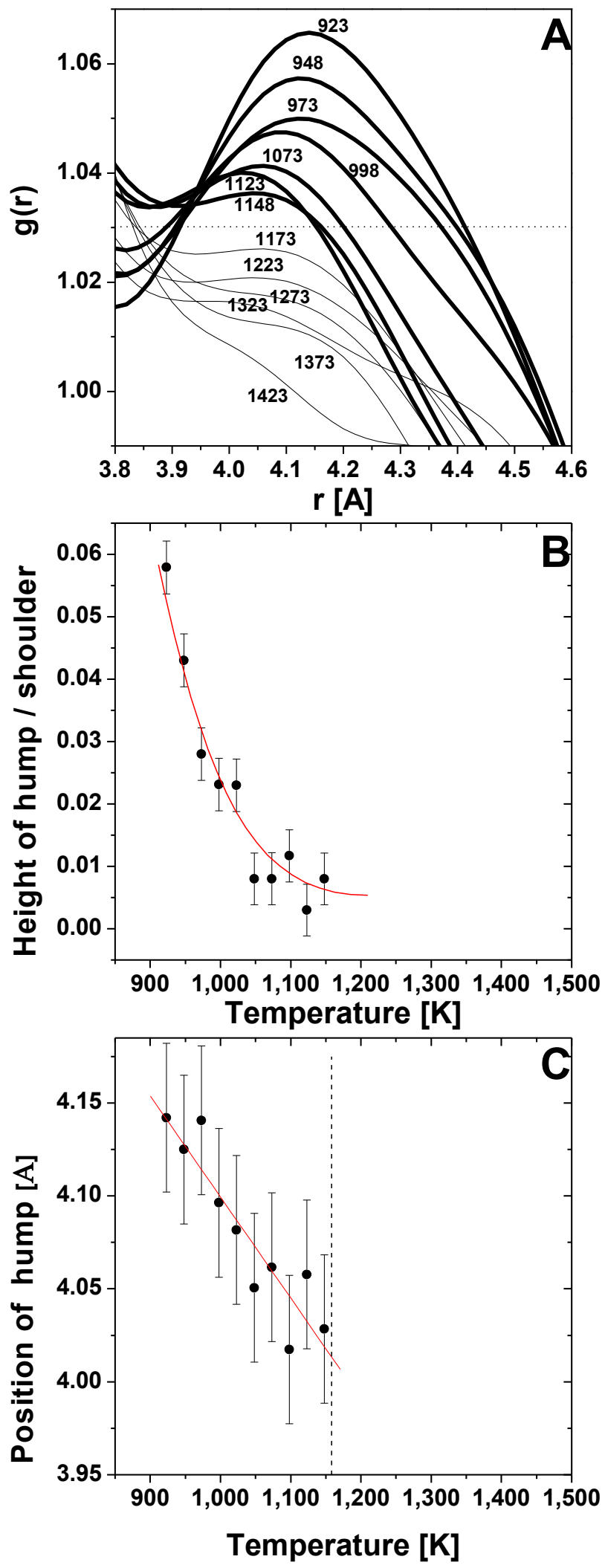

Figure 3: Analysis of the hump region in the radial distribution function (g(r)) in liquid antimony. (A) - Temperature dependence of g(r) in the hump region. (B) Hump height with respect to the preceding minimum in g(r) (C) - Hump position as a function of temperature. The transition temperature of ca. 1160K is marked by a dashed line. Note that above this temperature the hump is undefined.

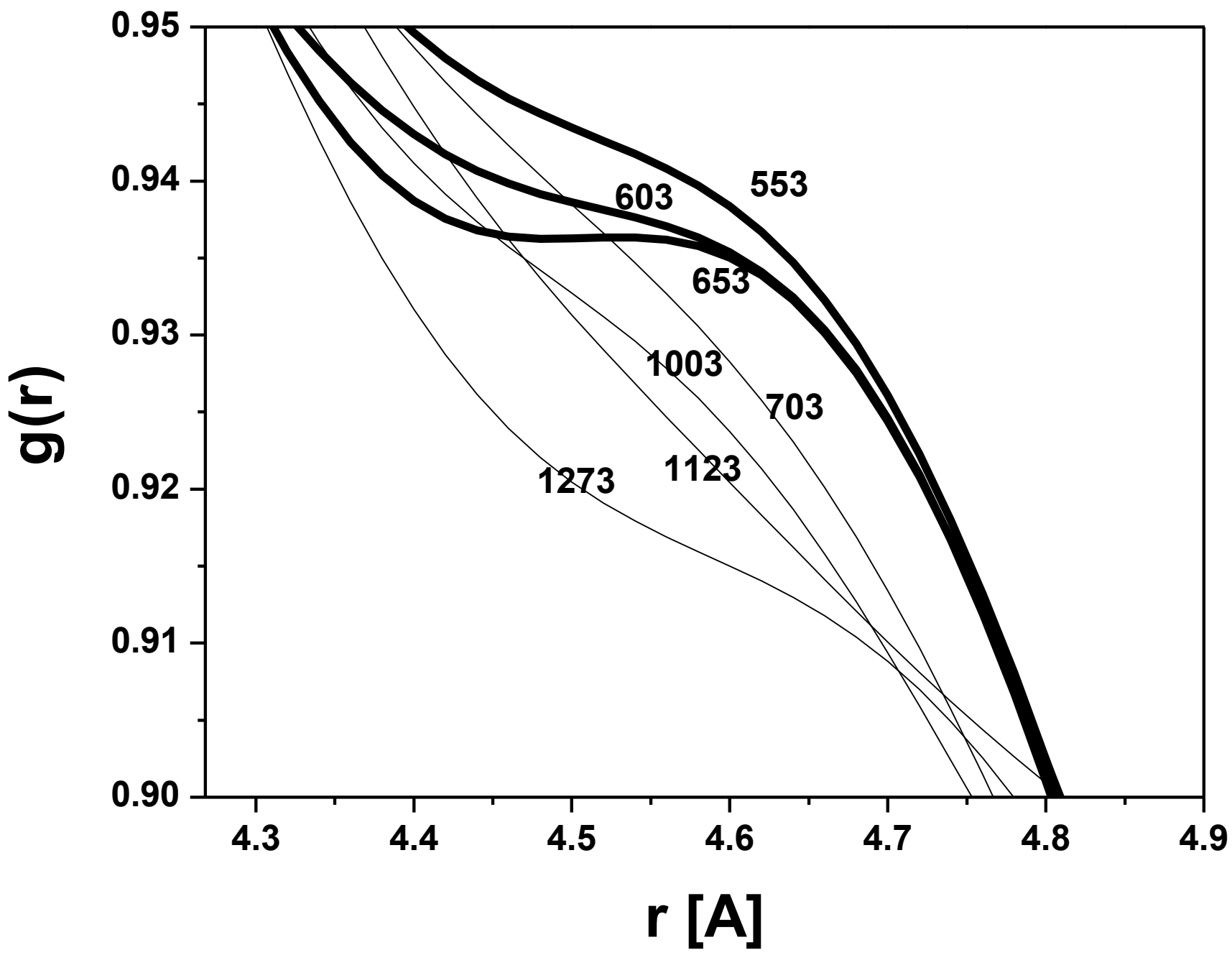

Figure 4: Temperature dependence of the shoulder region in the radial distribution function (g(r)) in liquid bismuth.

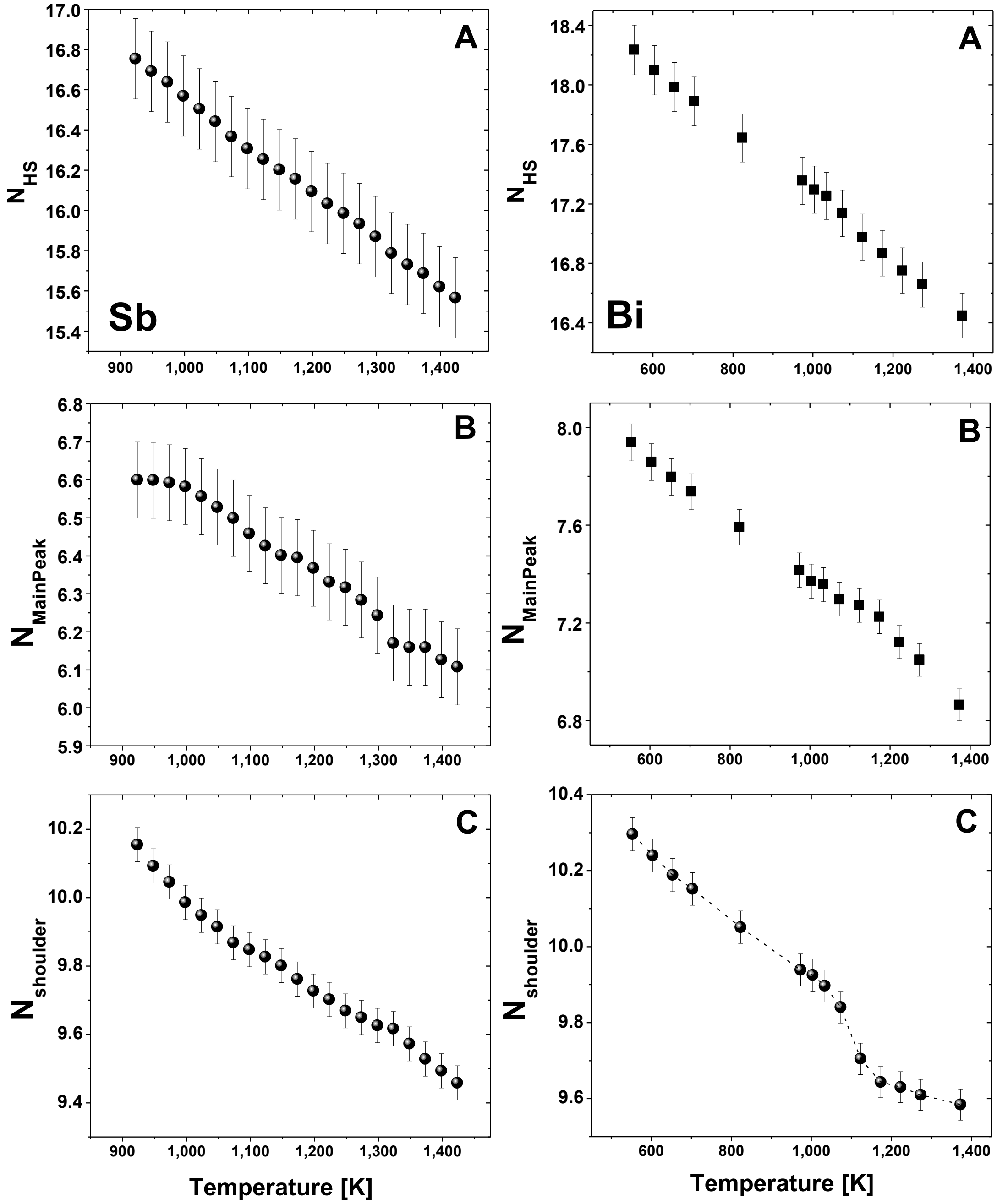

Figure 5: Coordination number of the first HS peak (A), main peak (B) and shoulder (C) as a function of temperature in liquid antimony and bismuth.

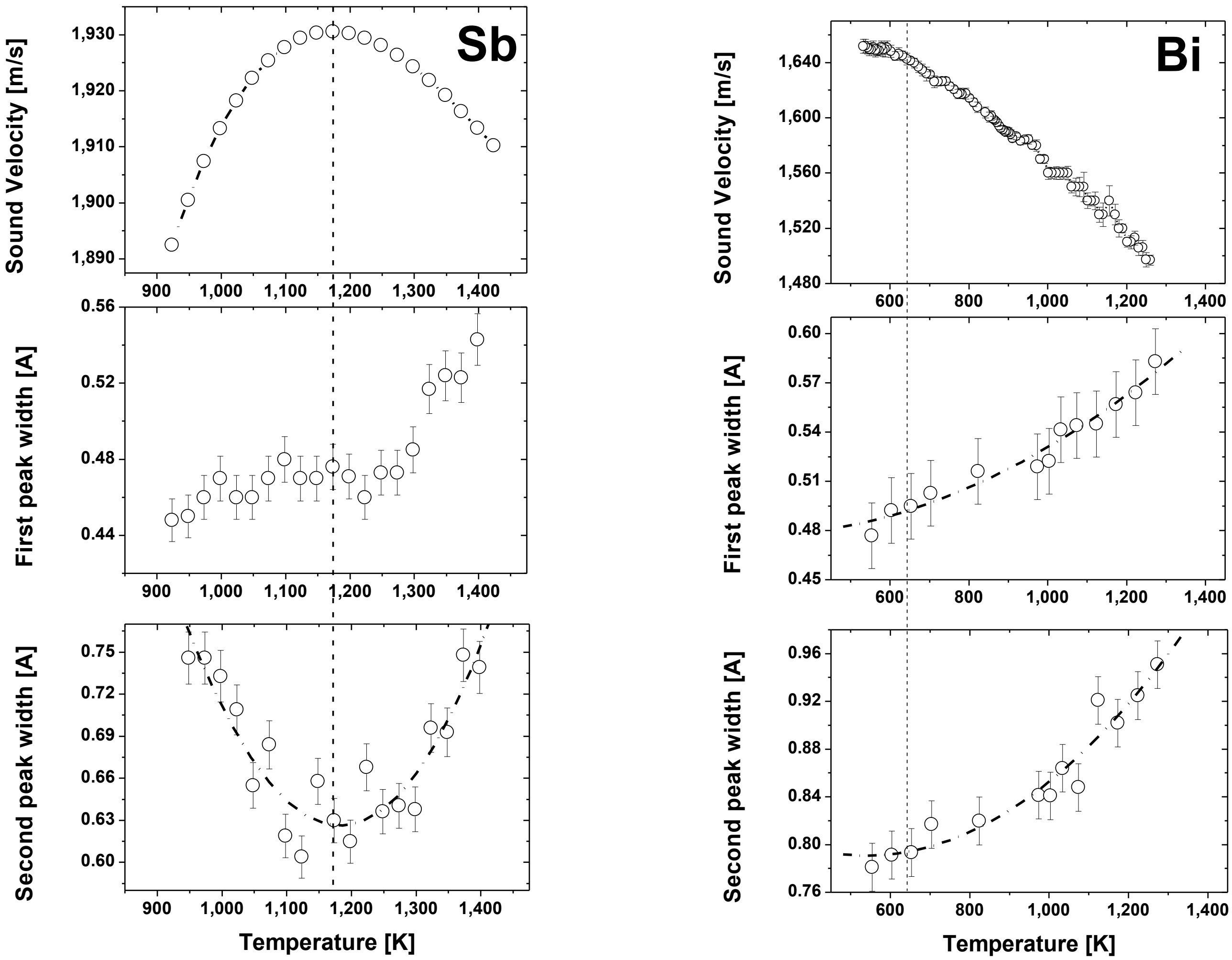

Figure 6: Temperature dependence of the sound velocity and of the FWHM of the first and second peaks in the radial distribution function of l-Sb (left) and l-Bi (right).

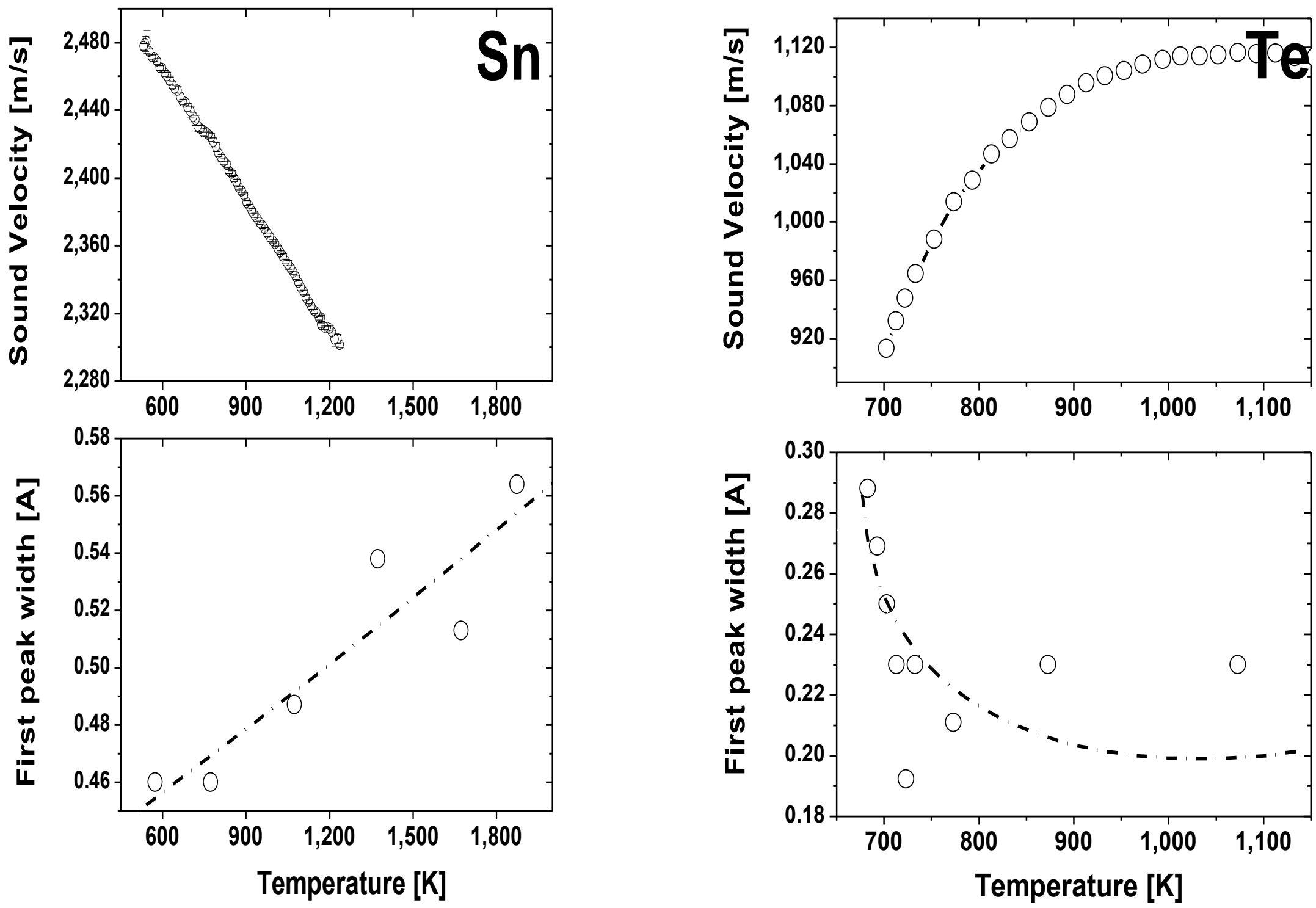

Figure 7: Temperature dependence of the sound velocity and of the FWHM of the first peak in the radial distribution function of l-Sn [6,20] and l-Te [32,54].